\documentstyle[epsfig]{mn2e}
\begin{document}
\input BoxedEPS.tex
\newcommand{\aip}{{\small ${\cal AIPS}$}}
\newcommand{\gtsim}{\mbox{{\raisebox{-0.4ex}{$\stackrel{>}{{\scriptstyle\sim}}
$}}}}
\newcommand{\ltsim}{\mbox{{\raisebox{-0.4ex}{$\stackrel{<}{{\scriptstyle\sim}}
$}}}}
\newcommand{\s}{$\stackrel{\rm s}{.}$}
\newcommand{\h}{$^{\rm h}$}
\newcommand{\m}{$^{\rm m}$}
\newcommand{\pp}{$\stackrel{\prime\prime}{.}$}
\newcommand{\de}{$^{\circ}$}
\newcommand{\p}{$^{\prime}$}
\newcommand{\arc}{$^{\prime\prime}$}
\newcommand{\marc}{^{\prime\prime}}
\newcommand{\rs}{{\em $r_s$}}
\newcommand{\DPM}{{\em DPM}}
\newcommand{\alf}{{\displaystyle\biggl({\nu_{\rm h} \over \nu_{\rm l}}\biggr)^{\alpha}} }

\newcommand{\figstart}[1]
    { \begin{figure}[htb]
      \begin{picture}(0,#1) }
\newcommand{\figend}[4]
    { \end{picture}
      \special{#1}
      \caption[#2]{#3}
      \label{#4}
      \end{figure} }
\newcommand{\fig}[5]
    { \figstart{#1}
      \figend{#2}{#3}{#4}{#5} }
\newcommand{\bHS}{\beta_{\mbox{\scriptsize HS}}}
\newcommand{\bBF}{\beta_{\mbox{\scriptsize BF}}}
\newcommand{\nT}{\nu_{\mbox{\scriptsize T}}}
\newcommand{\et}{E_{\mbox{\scriptsize T}}}
\newcommand{\nTn}{\nu_{\mbox{\scriptsize Tn}}}
\newcommand{\nTf}{\nu_{\mbox{\scriptsize Tf}}}
\newcommand{\tn}{\tau_{x\mbox{\scriptsize n}}}
\newcommand{\tf}{\tau_{x\mbox{\scriptsize f}}}
\newcommand{\xn}{x_{\mbox{\scriptsize n}}}
\newcommand{\xf}{x_{\mbox{\scriptsize f}}}
\newcommand{\yn}{y_{\mbox{\scriptsize n}}}
\newcommand{\yf}{y_{\mbox{\scriptsize f}}}
\newcommand{\lln}{l_{\mbox{\scriptsize n}}}
\newcommand{\llf}{l_{\mbox{\scriptsize f}}}
\newcommand{\Dn}{f(\Delta_{\mbox{\scriptsize n}})}
\newcommand{\Df}{f(\Delta_{\mbox{\scriptsize f}})}
\newcommand{\B}{\mbox{$B$}}
\newcommand{\Bo}{\mbox{$B$}_{0}}

\SetRokickiEPSFSpecial
\HideDisplacementBoxes

%%%%%%%%%%%%%%% MODIFICATION HISTORY %%%%%%%%%%%%%%%%%%%%
%%%%%
%%%%%  
%%%%%%%%%%%%%%%%%%%%%%%%%%%%%%%%%%%%%%%%%%%%%%%%%%%%%%%%%

\title[AGN models and the IR spectrum of IRAS F10214+4724]
{Active galactic nucleus torus models and the puzzling infrared spectrum of IRAS F10214+4724}
  \author[Efstathiou et al.]
{\parbox{\textwidth}{A.~Efstathiou,$^{1}$\thanks{E-mail: \texttt{a.efstathiou@euc.ac.cy}}
N.~Christopher,$^2$
A.~Verma,$^2$
R.~Siebenmorgen$^3$
}\vspace{0.4cm}\\
\parbox{\textwidth}{
$^1$School of Sciences, European University Cyprus, 
Diogenes Street, Engomi, 1516 Nicosia, Cyprus.\\
$^2$Oxford Astrophysics, Denys Wilkinson Building, University of Oxford,
Keble Road, Oxford OX1 3RH, UK\\
$^3$European Southern Observatory, Karl-Schwarzschildstrasse 2, 85748 Garching b. M\"unchen, Germany\\
}}
\maketitle

\begin{abstract}

 We present a revised model for the infrared emission of the hyperluminous
infrared galaxy IRAS F10214+4724 which takes into account recent photometric
 data from {\em Spitzer} and {\em Herschel} that sample the peak of its spectral
 energy distribution. We first present and discuss a grid of smooth active galactic
 nucleus (AGN) torus models computed with the method of Efstathiou \& Rowan-Robinson
 and demonstrate that the
 combination of these models and the starburst models of Efstathiou and coworkers,
 while able to give an excellent fit to the average spectrum of Seyfert 2s  and
 spectra of individual type 2 quasars measured by {\em Spitzer}, fails to match the spectral
 energy distribution of IRAS F10214+4724. This is mainly due to the fact that the
 $\nu S_{\nu}$ distribution of the galaxy falls very steeply with increasing frequency
 (a characteristic that is usually indicative of heavy absorption by dust) but shows
 a silicate feature in emission. Such emission features
 are not expected in sources with optical/near-infrared type 2 AGN spectral signatures.
 The {\em Herschel} data show that there is more power emitted in the rest-frame
 20-50$\mu m$ wavelength range compared with the model presented by Efstathiou 
 which assumes three components of emission: an edge-on torus, clouds (at a temperature
 of 610 and 200K) that are associated with the narrow-line region (NLR) and a highly 
obscured starburst that dominates in the submillimetre. We present a revised version 
of that model that assumes an additional component of emission which we associate with
 NLR clouds at a temperature of 100K. The 100K dust component could also be explained
 by a highly obscured hot starburst. The model suggests that the NLR of IRAS F10214+4724
 has an unusually high covering factor ($\geq 17\%$) or more likely the magnification
 of the emission from the NLR clouds is significantly higher than that of the emission
 from the torus.

\end{abstract}

%\large

\section{Introduction}
\label{}

 The discovery paper of IRAS F10214+4724 at a redshift of 2.285
(Rowan-Robinson et al. 1991) was entitled `An IRAS galaxy with huge
 luminosity - hidden quasar or protogalaxy?'. It soon became evident 
that gravitational lensing of a factor between 10 and 100 was responsible 
for the enormous luminosity that appears to exceed $10^{14} L_\odot$
 (Broadhurst \& Lehar 1995; Graham \& Liu 1995; Serjeant et al. 1995;
 Eisenhardt et al. 1996). Until recently, however, it was not clear 
whether the dominant energy source was a quasar or a starburst. The
 rest-frame optical/ultraviolet (optical/UV) spectrum showed 
characteristics similar to that
 of a Seyfert 2 (Elston et al. 1994; Lawrence et al. 1993; Goodrich 
et al. 1996) but the galaxy is also bright in the submillimetre
 suggesting the presence of an obscured starburst.

 Gravitational lensing can be turned into an advantage and can help us to
 study in more detail an object that is at a similar redshift to large numbers
 of ultraluminous and hyperluminous infrared galaxies (ULIRGs and HLIRGs 
with luminosities exceeding $10^{12}$ and $10^{13}L_\odot$ respectively)
 that have been discovered in large surveys at far-IR and submillimetre 
wavelengths (e.g. PEP: Lutz et al. 2011, HerMES: Oliver et al. 2012). Recent
 results from such surveys suggest that high redshift ULIRGs are  colder
 and less obscured than local ULIRGs (Rowan-Robinson et al. 2010, Elbaz et al.
 2011) as suggested by radiative transfer modelling of submillimetre galaxies
 (Efstathiou \& Rowan-Robinson 2003, Efstathiou \& Siebenmorgen 2009). 

 A number of models for the IR spectral energy distribution (SED) of
 IRAS F10214+4724 were developed in the early- to mid- 1990s but they were
inconclusive mainly due to the limited wavelength coverage. Rowan-Robinson et
 al. (1993) showed that good fits to the SED  could be obtained either with
 a starburst or a dust embedded quasar. Green \& Rowan-Robinson (1996) suggested
 that a combination of an active galactic nucleus (AGN) torus model and a 
 starburst was necessary to explain the SED. Granato et al. (1996) presented
 a fit to the SED of IRAS F10214+4724 solely with an embedded quasar.

 Mid-IR spectrophotometry has been shown to be a powerful means of 
differentiating between a starburst and an AGN
 origin of the IR emission (Roche et al. 1991 and references therin,
 Genzel et al. 1998, Verma et al. 2005).
Starburst galaxies show mid-IR emission features that are attributed
 to polycyclic aromatic hydrocarbon (PAH) molecules. These features 
are mostly absent from the spectra of quasars (QSOs)
 but in general AGNs may show mid-IR emission features with smaller
 equivalent widths due to the mixing of the starburst and AGN torus emission 
(e.g. Rowan-Robinson \& Efstathiou 2009). The spectra of quasars also show
 emission features at 10$\mu m$ whereas Seyfert 2s show absorption features
 due to silicate dust. 

In contrast, the mid-IR spectrum of IRAS F10214+4724 obtained by Teplitz
 et al. (2006) using the IRS onboard {\em Spitzer} was very puzzling as it
 showed an emission feature at rest frame 10$\mu m$ instead of the expected
 absorption feature given the type 2 nature of the AGN suggested by optical
 observations. Efstathiou (2006; hereafter Paper I) proposed that the SED of
 IRAS F10214+4724 could be the combination of three components of emission:
 discrete clouds with a covering factor of 17\% which were associated with the
 narrow-line region, an AGN torus viewed edge-on and a highly obscured starburst.
 We also noted in Paper I, however, that because the emission of the torus and the
  NLR clouds arises from different regions the magnification of each
 component may be different. 

 In this paper, we develop a new model for the IR emission of  IRAS F10214+4724
 which takes into account new photometric data from {\em Spitzer} and {\em Herschel}
 that cover the wavelength range 70-170$\mu m$ and therefore complete its SED. In
 Section 2, we present the new IRAC and MIPS data. In Section 3, we discuss a recently
 computed grid of tapered disc models computed with the method of Efstathiou \&
 Rowan-Robinson (1995) and a grid of starburst models computed with the method of 
Efstathiou, Rowan-Robinson \& Siebenmorgen (2000; hereafter ERRS00). We then show that while the combination
 of the tapered disc and starburst models gives an excellent fit to the average 
spectrum of Seyfert 2 galaxies presented by Hao et al. (2007)  and the spectra of 
type 2 QSOs (QSO2s) observed with {\em Spitzer}/IRS, it can not explain the SED of IRAS F10214+4724.
 In Section 5, we present our revised model for IRAS F10214+4724 and in Section 6 we discuss
 our results. A flat Universe is assumed with $\Lambda =0.73$ and H$_0$=71km/s/Mpc.

\section{{\em Spitzer} and {\em Herschel} observations}

We complemented the data from Paper I with imaging and photometry data from the {\em Spitzer
 Space Telscope} ({\em Spitzer}; Werner et al. 2004)  Heritage Archive and data taken with the
 Photodector Array Camera and Spectrometer (PACS; Poglitch et al. 2010) onboard the {\em Herschel
 Space Observatory} ({\em Herschel}; Pilbratt et al. 2010).

IRAS F10214+4724 was observed with the Infrared Array Camera (IRAC) at 3.6, 4.5,
5.8 and 8.0 $\mu$m as part of the PRME-MISC programme (AOR: 4405248). Multiband Imaging
 Photometer (MIPS; Rieke et al. 2004) images were also obtained as part of the ULIRGS-GTO2
 programme (AOR: 17537280) in the 70 and 160$\mu$m bands.

Imaging data from all mentioned {\em Spitzer} observations were used in this work,
 except for the MIPS 160$\mu$m, as the mosaic was of insufficient size to 
accurately determine the background, which contained significant structure.
 As Sturm et al. (2010) made a flux continuum measurement at 170$\mu m$, the lack
 of the 160$\mu m$ constraint does not affect this work.

The Basic Calibrated images (BCD) were obtained from the Spitzer Heritage 
Archive, then reduced and mosaicked using the MOPEX software (version 18.4.9).

These images were corrected for geometric distortions and interpolated on to an output
 grid, defined by the spatial boundaries of the BCD images included in the final mosaic. 
The background between overlapping frames was matched and bad pixels were masked out and
 these pixels reinterpolated. The average of each group of input pixel values was then used to
 create the final mosaic image.

The IRAC BCD images were produced by the S18.18.0 version of the  reduction software and
 calibrated according to Cohen et al. (2003). The MIPS 70$\mu$m BCDs were produced by version
 S18.12.0 of the reduction software and calibrated according to Gordon et al. (2007). 

Aperture photometry was then conducted to calculate the flux of IRAS F10214+4724
in the 3.6, 4.5, 5.8, 8.0 and 70$\mu$m {\em Spitzer} wavebands. An aperture radius of 3.6 arcsec
 was adopted for the IRAC mosaics and aperture corrections from the IRAC instrument
 handbook were applied for a background radius of 14.4-24 arcsec. No colour corrections were 
applied. A calibration uncertainty of 3\% was implemented, 
as discussed in Reach et al. (2005). Possible contributions to the IRAC fluxes from the
 foreground lensing source were not estimated.

An aperture radius of 16 arcsec was used for the MIPS 70$\mu$m mosaic and  the photometry was 
corrected for using aperture corrections based on the best-fitting MIPS point spread function from Gordon
 et al. (2008). A colour correction of 0.98 was determined according to the measured SED
 of IRAS F10214+4724 approximated by a power-law slope across the MIPS bands (Stansberry et al.
 2007). An additional calibration uncertainty of 5\% was also applied, as described in Gordon
 et al. (2007). Aperture sizes and aperture corrections as well as source fluxes and the associated
 error per {\em Spitzer} band are listed in Table 1.

Fluxes at 85, 108 and 170$\mu$m (observed wavelength) were obtained by PACS as part
 of the SHINING programme and are published in Sturm et al. (2010). These continuum fluxes were
 obtained from the line spectra of [OIV]26$\mu$m, [SIII]33$\mu$m and [OIII]52$\mu$m. The PACS
 fluxes are also given in Table 1.

\begin{table*}
\caption{Summary of the new {\em Spitzer}, {\em Herschel}/PACS and other data for IRAS F10214+4724 which
 were used in the radiative transfer modelling. The listed error is the combination of the flux
 error and the adopted calibration uncertainty (3$\%$ of the source flux for IRAC bands and 5\%
 for MIPS 70$\mu$m). References for the data (1) This work (2) Sturm et al. (2010) (3) IRAS (4)
 Benford et al. (1999) (5) Rowan-Robinson et al. (1993) 
}
\begin{center}
\begin{tabular}{c c c c}
Wavelength ($\mu$m)   & Flux (mJy)&  Error (mJy) & Reference \\
\hline
\hline
Spitzer&       &       & \\
\hline
3.6    & 0.32	&  0.06 &   1\\
4.5    & 0.33 	&  0.06 &   1\\
5.8    & 0.43  &  0.07 &   1\\
8.0    & 0.96  &  0.10 &   1\\
70.0   & 301   &   29  &   1\\
       &       &      &     \\
{\em Herschel}/PACS   &       &      &     \\
\hline
85     & 330   &  100 & 2\\
108    & 378   &  115 & 2\\
170	 & 445   &  130 & 2 \\
       &       &      & \\
Other data  & & &\\
\hline
60    & 205  & 45  & 3\\
350   & 383  & 51  & 4\\
450   & 273  & 45  & 5\\
800   & 50   & 5   & 5\\
1100  & 24   & 5   & 5\\
\hline
\end{tabular}
\end{center}
\end{table*}

\begin{figure*}
\epsfig{file=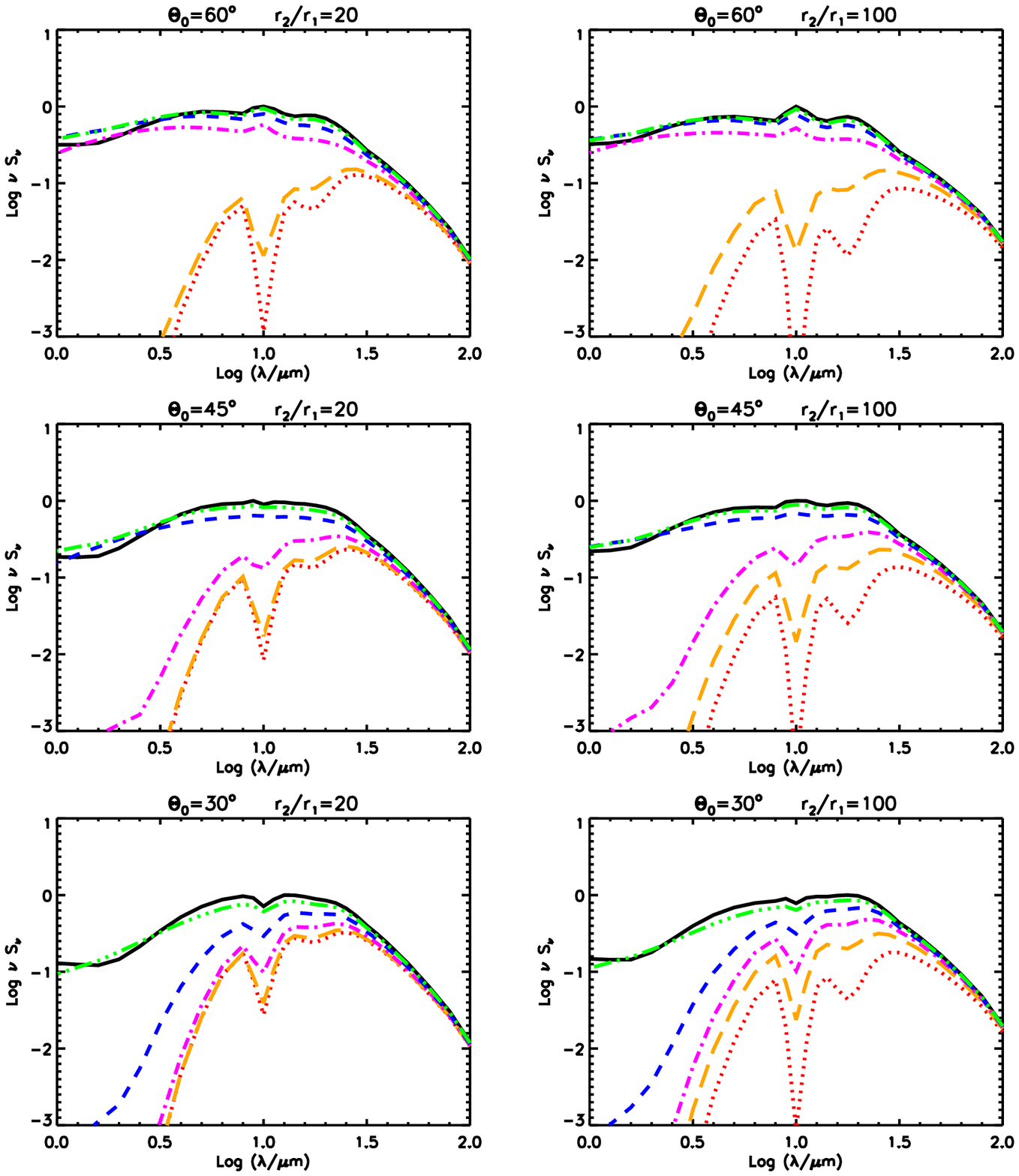, angle=0, width=15cm}
\caption{
Predicted SEDs of tapered discs for a range of half-opening angles
 of the torus $\Theta_0$ and inclinations, and two different values for the ratio of the outer to
 inner disc radius $r_2/r_1$. All models assume an equatorial optical depth at 1000\AA~ of 750
 and a density distribution that falls off as $r^{-1}$. We use the following code for the 
inclinations in all panels: red dotted line -- $90^o$, orange dashed line -- $75^o$, magenta dot
 dashed line -- $60^o$, blue dashed line -- $45^o$, green triple dot dashed line -- $30^o$, black
 solid line -- $0^o$, }
\end{figure*}

\section{Radiative transfer models}

\subsection{AGN torus models}

When the first models for the emission of the torus in AGNs were developed (Pier \& Krolik 1992,
 Granato \& Danese 1994, Efstathiou \& Rowan-Robinson 1995), the effort of these authors was mainly
 guided by the IRAS data and ground-based spectrophotometry in the 8-13$\mu m$ window by Roche,
 Aitken and collaborators (Roche et al. 1991, and references therein). A spectrum of the nucleus
 of NGC 1068 from 2.2 to 34$\mu m$ was also available (Rieke \& Low 1975). The spectrophotometry showed
 moderate silicate absorption in Seyfert 2s and featureless spectra in Seyfert 1s. The absence
 of silicate emission in Seyfert 1s was very puzzling as in the context of the AGN unified scheme
 (e.g. Antonucci 1993) in Seyfert 1s we should be seeing the hot dust in the inner part of the
 obscuring torus directly and this should give rise to a silicate emission feature (Efstathiou
 \& Rowan-Robinson 1991, Rowan-Robinson et al. 1993). Two solutions for this problem were proposed:
 Granato \& Danese (1994) suggested that silicate dust is destroyed in the inner part of the torus
 by shocks. Pier \& Krolik (1992) and Efstathiou \& Rowan-Robinson (1995) made the torus so optically
 thick that it was optically thick at 10$\mu m$ even when it was viewed face-on. Pier \& Krolik 
explored the case of cylinders of constant density, whereas Efstathiou \& Rowan-Robinson that of
 tapered discs with a density distribution following $r^{-1}$. The thickness of tapered discs
 increases with distance from the central source in the inner part of the disc but tapers off
 to a constant value in the outer disc. Because of the differences in geometry and density
 distribution the cylindrical and tapered disc models make significantly different predictions
 about the emergent spectral energy distributions.  The tapered discs of Efstathiou \& Rowan-Robinson,
 in combination with the starburst models of ERRS00 
 have been quite successful in fitting the SEDs of a number of AGNs even
 in cases where mid-IR spectroscopy was available (Alexander et al. 1999, Ruiz et al. 2001,
  Farrah et al. 2002, Verma et al. 2002, Farrah et al. 2003, Efstathiou \& Siebenmorgen 2005, Farrah
 et al. 2012).

More recently there has been interest in clumpy torus models (Nenkova et al. 2002, Dullemond
 \& van Bemmel 2005, H\"onig et al. 2006, Nenkova et al. 2008, Schartmann et al. 2008, Heymann \& Siebenmorgen 2012,
 Stalevski et al. 2012) for two main reasons: first,  these authors were attempting
 to explain the absence of silicate emission features in type 1 AGNs with clumpiness following an
 original suggestion by Rowan-Robinson (1995) and, secondly,  theoretically it is
 predicted that the dust in the torus should be clumpy. The irony is that in 2005 {\em Spitzer}
 observations showed  emission features in quasars and weak absorption features in Seyfert 1 
galaxies (e.g.  Hao et al. 2005, 2007, Siebenmorgen et al. 2005, Spoon et al. 2007). It is currently
 not clear whether these emission features are emitted by the torus or by NLR clouds
 as suggested in Paper I and by Schweitzer et al (2008). It is also not clear whether clumping of
 the dust in the torus is needed at all for explaining the SEDs of AGNs
 (see also Feltre et al. 2012). 

In this paper we use a grid of tapered disc models computed with the method of Efstathiou \&
 Rowan-Robinson (1995). In this grid of models we consider five discrete values for the equatorial
 1000\AA~ optical depth (250, 500, 750, 1000, 1250), three  values for the ratio of the outer to inner
 disc radius $r_2/r_1$ (20, 60, 100) and four values for the half-opening angle of the torus $\Theta_0$
 (30$^o$, 45$^o$, 60$^o$ and 75$^o$; $\Theta_1$ as defined by Efstathiou \& Rowan-Robinson is equal
 to $90 - \Theta_0$). The spectra are computed for 37 or 74 orientations which are equally spaced
 in the range 0 to $\pi/2$.  The method of solution of the axially symmetric radiative transfer
 problem has been tested extensively and has been shown to give results that are in very good
 agreement with the benchmark models of Pascucci et al. (2004). The accuracy of the results is
 also ensured by the very good satisfaction of the flux constancy condition even in the most
 optically thick multi-grain models. 

 In Fig. 1 we show the predicted SEDs from six models that illustrate the range of spectra that
 can be obtained from these models. In these models, we fix the equatorial optical depth at 1000\AA~
 at 750 and vary the torus half-opening angle $\Theta_0$ and the ratio of the outer to inner torus radius
 $r_2/r_1$. We plot the predicted spectra for the same inclinations in all panels for easy comparison.
 Most of the effects seen in Fig. 1 have been discussed in Efstathiou \& Rowan-Robinson (1995) but
 it is useful to make some additional comments here now that we have the benefit of a larger regular
 grid of models. First of all, it is clear that in the face-on case we can have flat 10$\mu m$ spectra
 as well as spectra with emission and absorption features. Secondly, for a fixed outer to inner disc
 radius, the anisotropy of the emission decreases with decreasing torus half-opening angle. Tori with
 a smaller half-opening angle emit more isotropically as they deviate less from spherical symmetry. 
Finally for a fixed opening angle, tori that are more extended (or, in other words, have a larger $r_2/r_1$)
 show more anisotropy in the emission. We also note that the anisotropy of the emission increases with
 increasing equatorial optical depth.  The grid of tapered disc models, as well as the grid of starburst
 models discussed below, are available from AE on request (a.efstathiou@euc.ac.cy). 

\subsection{Starburst models}

Models for the IR emission of starburst galaxies have been presented by
Rowan-Robinson \& Crawford (1989), Rowan-Robinson \& Efstathiou (1993), Kr\"ugel
 \& Siebenmorgen (1994), Silva et al. (1998), ERRS00, Takagi et al. (2003), Dopita
 et al. (2005), Siebenmorgen \& Kr\"ugel (2007) and Groves et al. (2008).

The model of ERRS00,   which has
 been updated by Efstathiou \& Siebenmorgen (2009), incorporates the stellar population synthesis
 model of Bruzual \& Charlot (1993, 2003) and involves detailed radiative transfer that takes into
 account absorption, emission and multiple scattering from large classical grains as well as the
 effect of small transiently heated grains and PAH molecules. The emission of small grains and PAHs
 is calculated with the method of Siebenmorgen \& Kr\"ugel (1992). One distinct feature of the model
 is that it also incorporates a simple recipe for the evolution of the giant molecular clouds that
 constitute the starburst. The model predicts that by about 10 Myr after star formation, the expansion
 of the H$_{II}$ region leads to the formation of a cold narrow shell of gas and dust. This naturally
 explains why the mid-IR spectra of starburst galaxies are dominated by the PAH emission
 and not by the emission of hot dust. The model predicts that as the molecular clouds  evolve, their
 emission shifts to longer wavelengths and show stronger PAH features. 

Another effect of ageing is that the clouds get more optically thin and therefore the silicate
 features get shallower. Once the spectra of molecular clouds at different stages in their
 evolution have been computed, the spectrum of a starburst at different ages can be computed
 by convolving the sequence of spectra with a star formation history. ERRS00 showed that the
 IRAS colours of starburst galaxies can be explained by a sequence of models whose star formation
 rate decays exponentially with an e-folding time $\tau $ of 20Myrs. The other model parameters
 assumed by ERRS00 (average density  $n_{av}$, star formation efficiency $\eta$ and giant molecular
 cloud mass $M_{GMC}$) were found to be in good agreement with other indicators for the starburst
 galaxy M82.  As it is well known (e.g. Rowan-Robinson \& Efstathiou 1993), ULIRGs such as Arp 220
 are up to a factor of a few more optically thick than starbursts like M82
 and NGC 1068. Verma et al. (2002) and Farrah et al. (2002) showed that the starburst models of ERRS00
 in combination with the tapered discs of Efstathiou \& Rowan-Robinson (1995) could explain the SEDs
 of HLIRGs. Farrah et al. (2003) showed that the same combination of models
 provided good fits to the SEDs of 41 ULIRGs.  Rowan-Robinson \& Efstathiou
 (2009) found that the same models can explain the full range of spectra measured by the infrared
 spectrograph (IRS) onboard {\em Spitzer} (Spoon et al. 2007). One interesting prediction of the
 ERRS00 model is that young starbursts show strong near-IR continua, small 6.2$\mu m$ equivalent
 widths and deep silicate absorption features and can therefore explain the galaxies in class 3A of
 Spoon et al. (2007). Recently Rowan-Robinson et al. (2010) found a large number of young starbursts
 in the HerMES project.

 In this paper we use a grid of starburst models that have been computed with the method of ERRS00
 but with a revised dust model (Efstathiou \& Siebenmorgen 2009)
and revised stellar population synthesis models (Bruzual \& Charlot 2003).
 We first vary $n_{av}$, $\eta$ and $M_{GMC}$ to give three discrete values of $\tau_V$
 (50, 75 and 100) that cover the range of optical depth suggested by observations. We then
 compute a grid of starburst models in which we vary the age of the starburst $t_*$ from 0
 to 70Myrs in steps of 5Myr, the e-folding time of the exponentially decaying star-formation
 rate $\tau $ from 10 to 40Myr in steps of 10Myr and the time $t_m$ at which the covering factor
 of dust reduces from 100 to 50\% from 10 to 40Myr also in steps of 10Myr. This gives us a total
 of 720 starburst models. 

\subsection{Comparison with the average IR spectrum of Seyfert 2s}

 We first explore whether the combination of the tapered disc and starburst models described
 above  can fit the average spectrum of Seyfert 2s presented by Hao et al. (2007). The average
 observed spectrum has been determined by first normalizing  the spectra at rest-frame 15$\mu m$
 and then averaging. The {\em Spitzer} spectrum  has been supplemented in the far-IR by
 determining the average IRAS fluxes of the sample of Seyfert 2s discussed by Lumsden et al. (2001)
 and normalizing them to the Spitzer spectrum at 12$\mu m$. The tapered disc models have also been
 averaged in an attempt to mimic the averaging of the observed spectra. For a given AGN torus model,
 the predicted spectra for the relevant inclinations have been normalized at 15$\mu m$ and then
 weighted over the viewing angle, i.e by carrying out the integral

 $$ \int_{i_1}^{\pi/2} \nu S_{\nu}(i)~ sin ~i~ di  $$

where $i_1$ is the angle that separates a direct view of the nucleus from an obscured view.

\begin{figure}
\epsfig{file=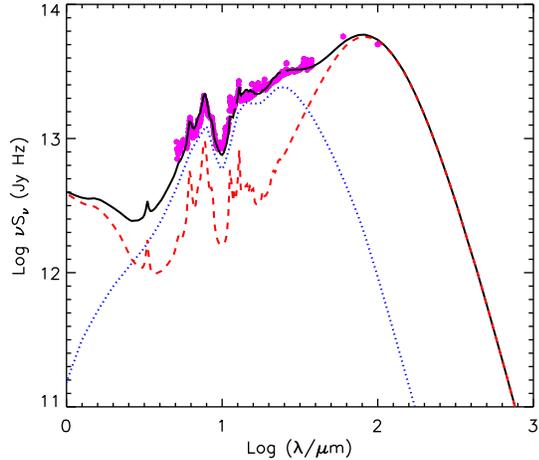, angle=0, width=8cm}
\caption{
 Fit to the average spectrum of Seyfert 2s by Hao et al. (2007). The blue dotted line gives
 the average emission from the AGN tapered disc model (see text) and the red dashed line the
 emission from the starburst/cirrus combination model (see text). The total emission is given
 by the solid line. The vertical scale is arbitrary.}
\end{figure}

\begin{figure}
\epsfig{file=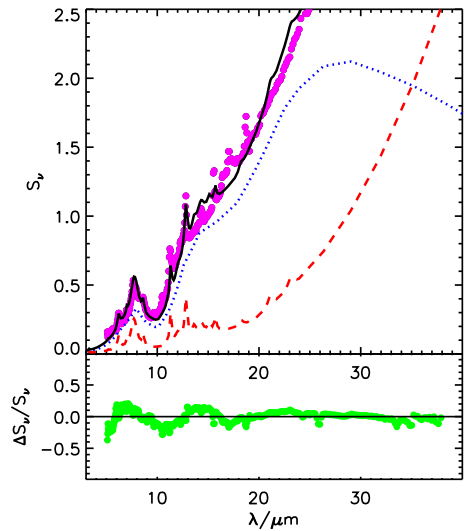, angle=0, width=10cm}
\caption{
 As Fig. 2 but showing the fit to the mid-IR spectrum
 on a linear scale in more detail. The vertical scale is arbitrary.
The lower panel shows the residuals $(S_{model} - S_{data})/S_{data}$.
}
\end{figure}

\begin{figure*}
\epsfig{file=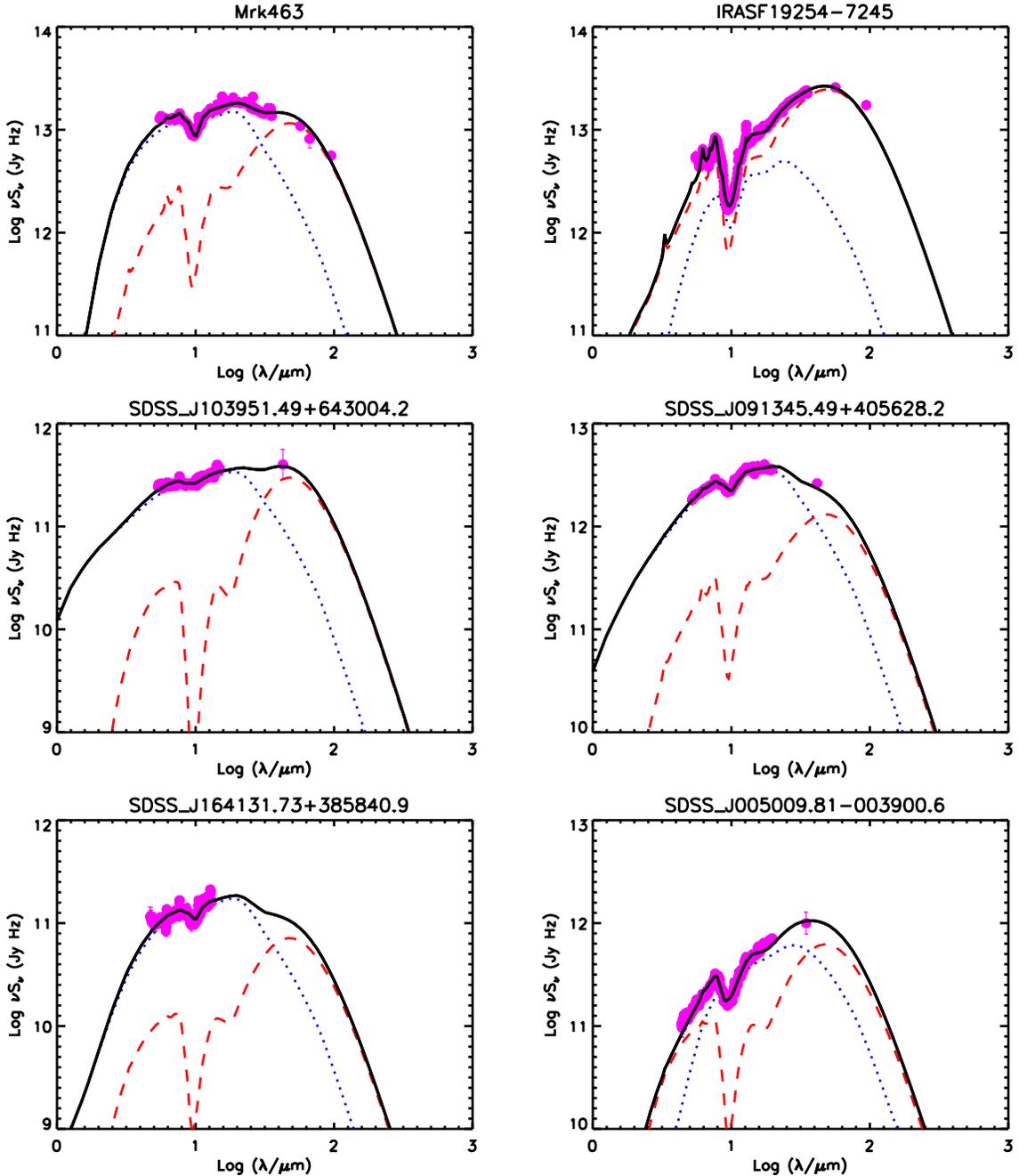, angle=0, width=16cm}
\caption{
 Best-fitting tapered disc (blue dotted) and starburst (red dashed) models for the QSOs discussed
 in section 4. The total emission is given by the black solid line. The best fit parameters
 and derived luminosities are given in Table 2.
Data from Hernan-Caballero \& Hatziminaoglou (2011) and IRAS.}
\end{figure*}

 We find that even the oldest of the starburst models considered here can not account for
 the average 60 and 100$\mu m$ fluxes of this sample which suggests that there is a significant
 contribution from cold galactic discs (cirrus) to the far-IR emission. A number of the 
galaxies in the Lumsden et al. sample do in fact show evidence for excess 100$\mu m$ emission
 which is indicative of cirrus. We therefore combine the starburst models with the cirrus models 
of Efstathiou \& Siebenmorgen (2009) which assume that the diffuse galactic dust is heated by a
 radiation field that has the spectrum of the interstellar radiation field in the solar
 neighbourhood but an intensity which is a factor of $\psi $ higher. We find that a cirrus
 model that has the same luminosity as the starburst and $\psi =10$ when combined with the
 tapered disc models provides a very good fit to the data.   
 The best fit to the Seyfert 2 average (shown in Figs 2 and 3) is provided by
a model that assumes a torus half-opening angle of 45$^o$, an equatorial 1000\AA~ optical
 depth of 750 and a ratio of the outer to inner torus radius of 20. The best-fitting starburst model
 assumes $\tau_V=100$, $t_*=40$Myr, $\tau =$10Myr and $t_m=$10Myr.

\begin{table*}
\caption{Best-fitting parameters and derived luminosities for the QSOs modelled with a combination
 of tapered disc and starburst models. $t_*$ is the age of the exponentially decaying starburst,
 $\tau_V$ is the initial optical depth of its molecular clouds, $r_2/r_1$ is the ratio of the outer
 to inner tapered disc radius, $\tau_{uv}^{eq}$  is its equatorial UV optical depth,  $i$ is
 the disc inclination and  $\Theta_0$ is its half-opening angle. $L_{AGN}$, $L_{SB}$ and $L_{tot}$ 
 are the 1-1000$\mu m$ AGN, starburst and total luminosities, respectively. The AGN luminosity
 is corrected by the anisotropy correction factor $A$ (Paper I).
}
\begin{center}
\begin{tabular}{l c c c r r c c c c c c}
\hline
\hline
Object name   & $z$  & $t_*$ & $\tau_V$ & $r_2/r_1$ & $\tau_{uv}^{eq}$  & $i$ & $\Theta_0$ & $A$ &
 log$~L_{AGN} \over L_\odot$ & log$~L_{SB} \over L_\odot$ & log$~L_{tot} \over L_\odot$ \\
  &    & ($10^7$yr) &  &  &   &  &  &  &  &   &  \\
\hline

Mrk463                   &  0.051 &   0.50 &  50 &  60 &  250 &  59$^o$ &  30$^o$  &  1.11 &  11.69 &  11.44 &  11.88\\
IRAS F19254-7245         &  0.062 &    1.50 &  50 &  60 &  750 &  64$^o$ &  45$^o$  &  1.89 &  11.52 &  11.96 & 12.10\\
SDSSJ103951.49+643004.2  &  0.402 &    0.00 &  75 &  60 &  500 &  51$^o$ &  45$^o$  &  0.86 &  11.96 &  11.74 &  12.16\\
SDSSJ091345.49+405628.2  &  0.441 &   0.50 &  50 & 100 &  750 &  36$^o$ &  30$^o$  &  0.61 &  12.87 &  12.56 &  13.04\\
SDSSJ164131.73+385840.9  &  0.596 &   0.00 &  50 & 100 &  500 &  40$^o$ &  30$^o$  &  0.75 &  11.94 &  11.59 &  12.10\\
SDSSJ005009.81-003900.6  &  0.728 &   0.00 &  50 &  20 & 1250 &  67$^o$ &  60$^o$  &  4.17 &  13.40 &  12.74 &  13.49\\

\hline
\end{tabular}
\end{center}
\end{table*}

\section{Comparison with the infrared spectra of QSO2s}

To assess how succesful are these models in explaining the IR spectra of more 
luminous AGN we have compared them with the SEDs of six QSO2s from the sample of
 Hernan-Caballero \& Hatziminaoglou (2011) which have good signal-to-noise ratio mid-IR
 IRS data and a range of spectral slope and apparent level of obscuration. We complement
 the IRS data with photometry at longer wavelengths where available.  For simplicity
 we assume $\tau =10$Myr and $t_m=t_*$. We find that the combination of tapered
 discs and starburst models presented here provide good fits to the data. The fits
 are shown in Fig. 4 and the best-fitting parameters and derived luminosities are given
 in Table 2.

The comparisons shown in this and the previous section show that existing models
 for starbursts, cirrus and smooth AGN tori with reasonable parameters provide 
good fits to the average spectrum of Seyfert 2 galaxies and spectra of individual
 QSO2s. The fit to the average spectrum of Seyfert 2s also shows that there is
 a significant contribution in the mid-IR spectra of this population by the
 starburst so any serious attempt to explain the {\em Spitzer} spectra of AGNs should
 consider a combination of AGN torus and starburst emission. 

\section{A revised model for IRAS F10214+4724}

 We now consider whether the same combination of models can fit the SED of IRAS
 F10214+4724. The broad-band data used in the fit are summarized in Table 1. In
 this analysis, we exclude the IRAC data as they require an additional starlight
 component to be fitted. In addition, the lensing galaxy may contribute as it is
 not spatially resolved from IRAS F10214+4724. We find that these models provide a very
 poor fit to the SED of  IRAS F10214+4724 and this is shown in Fig. 5. This is clearly
 due to the fact that the spectrum of IRAS F10214+4724 shows an unusual combination
 of a silicate emission feature and weak rest-frame near-IR continuum. This
 suggests that the only way we could fit this spectrum with a smooth torus viewed
 close to edge-on is if the torus was optically thin in the mid-IR and has
 a maximum dust temperature of 200-300K. Such models are not included in our grid
 of models all  of which assume a maximum dust temperature of 1000K and equatorial
 1000\AA~ optical depth in the range 250-1250. A model with a low torus optical
 depth is also inconsistent with the {\em Chandra} X-ray spectrum of F10214+4724 which
 suggests that if there is an AGN in this object it must be Compton-thick (Alexander et al. 2005).

 A question that naturally arises from the previous discussion is: is it possible to
 fit the spectrum of IRAS F10214+4724 with a clumpy torus model? Nikutta et al. (2009)
 presented a clumpy torus model for the SED of a type 2 quasar that shows a silicate
 feature in emission. In contrast to IRAS F10214+4724, this object shows a strong near-IR
 continuum as well. It is hard to see how even a clumpy torus with clouds as hot as 1000K
 viewed close to edge-on can produce a spectrum with a silicate emission feature and weak
 near-IR continuum. To investigate this further, we have compared the data with a combination
 of the clumpy torus models of Stalevski et al. (2012), which have been computed with a fully
 self-consistent radiative transfer code, and our starburst models. We find that the best fit,
 which is also shown in Fig. 5, is of similar quality to the fit obtained with the smooth
 torus models.

\begin{figure}
\epsfig{file=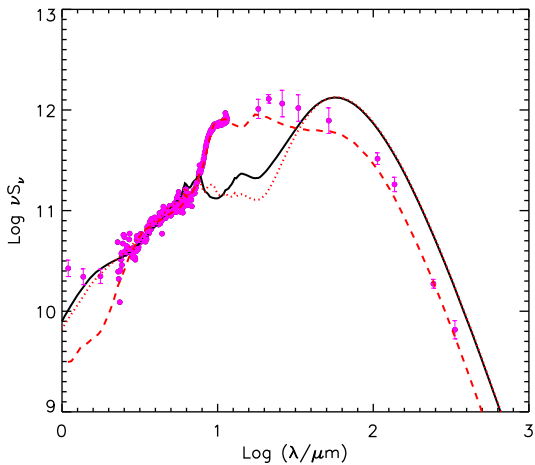, angle=0, width=8cm}
\caption{
The solid line shows the best fit to the SED of IRAS F10214+4724
 with a combination of the tapered disc models of Efstathiou \& Rowan-Robinson (1995) and
 the starburst models of Efstathiou \& Siebenmorgen (2009). Also shown as a dotted line (red)
 is the best fit to the data with the combination of the Stalevski et al (2012) clumpy torus
 models and the same starburst models. The dashed (red) line shows the fit to the data with
 the Paper I model. References for the data are given in Table 1.
}
\end{figure}

 The failure of torus emission alone to fit the mid-IR SED of IRAS F10214+4724 motivated us
 in Paper I to propose a three-component model: an edge-on torus, clouds associated with
 the NLR and a highly obscured starburst. The model assumed that the NLR
  clouds have a covering factor of 17\% and two discrete values for the dust
 temperature (610 and 200K). The eight-parameter fit with a model with two discrete temperatures
 for the NLR clouds is much more significant (reduced $\chi^2=2.5$) than a
 six-parameter fit with a model with only one temperature (reduced $\chi^2=11.8$). The
 parameters we varied are the temperature and luminosity of the two NLR components, the
 optical depth of the clouds, the luminosity of the starburst, the luminosity of the torus
 and its inclination.

We also found in Paper I that it is very difficult to fit the observed near- and mid-IR
 spectrum with a model that assumes that the NLR dust has a continuous distribution of
 dust temperature. In Fig. 8, we illustrate this by plotting the predicted spectrum of
 a shell of dust which has a ratio of the inner to outer radius of 0.1, a density distribution
 that follows $r^{-1}$ and a total UV optical depth of 10. We assume the same 
grain mixture as used for the calculation of the emission of discrete clouds. The
 dust temperature of the shell at its inner boundary is 610K and this reduces to  73 K at
 the outer radius. It is clear that such a model does not provide a good fit to the {\em Spitzer}
 spectrum.

 The new MIPS and PACS data suggest that there is about a factor of 2 more power in the rest-frame
 wavelength range 20-50$\mu m$ compared with the prediction of  the model presented in
 Paper I (see Fig. 5). Considering that the new data sample the peak of the SED of IRAS F10214+4724,
 this implies that the covering factor needs to increase even more from an already high value.
 We find that the new data can be fitted by assuming an additional component of NLR
 dust at a temperature of 100K. The new fit is shown in Figs 6 and 7. The combination
 of the anisotropy of the emission of the assumed torus model ($\Theta_0=60^o$, $r_2/r_1=20$
 and equatorial optical depth 1000) and the fact that the covering factor is so high  makes
 the torus emission negligible and it is not shown in this fit. Fig. 8 shows the individual
 contributions from the 610, 200 and 100K dust. The luminosity of the three components is $1.4$, 
$9.5$ and $7.1 \times 10^{13} L_\odot$, respectively, giving a total apparent AGN luminosity of
 $1.8 \times 10^{14} L_\odot$. The starburst model assumes the same parameters as the model
 that provides the best fit to the Seyfert 2 average except that it is 50\% more optically
 thick ($\tau_V = 150$). The starburst luminosity is predicted to be $6.6 \times 10^{13}
 L_\odot$ which is about 13\% lower than that found in Paper I.
   
\section{Discussion and Conclusions}

Before we discuss the proposed model for the IR emission of IRAS F10214+4724,
 we would like to discuss the possibility of an alternative interpretation for the
 additional component that is suggested by the {\em Herschel} data. We have explored 
the possibility that this additional component is a hot starburst. To this end, we 
searched the data base of starburst models of
Siebenmorgen \& Kr\"ugel (2007) for a model that peaks in the rest-frame wavelength
 range 20-40$\mu m$ and which is sufficiently optically thick so that its mid-IR
 emission does not affect the fit to the silicate emission feature which is still provided 
mainly by the 200K clouds.

 In the starburst model of ERRS00, the luminosity of the starburst and its radius are
 not free parameters. A more luminous starburst is assumed to consist of more molecular
 clouds and the model does not specify how these clouds are distributed in space. In the 
starburst model of Siebenmorgen \& Kr\"ugel (2007), the radius of the starburst (which is 
assumed to be spherical) and its luminosity are  free parameters. The model also assumes 
that in addition to the molecular clouds that are centrally illuminated by OB stars and 
which constitute the so-called hot spots,
there are other old bulge and massive stars that are obscured only by diffuse dust which 
is distributed uniformly in the starburst. The hot-spots and the free stars are distributed
 as $r^{-1.5}$ but the massive stars are confined to the central 350 pc.  By varying the 
luminosity and radius of the starburst it is therefore possible to heat the diffuse dust
 to a high temperature of about 100K.  

We find that a good fit to the complete SED can be obtained if we
 replace the 100 K component with a Siebenmorgen \& Kr\"ugel model that assumes a luminosity 
of $10^{14.2} L_\odot$, a radius of 1 kpc and  a visual extinction from the surface to the 
centre of the starburst of 119. The apparent luminosity of this component is about $7 \times
 10^{13} L_\odot$ so this implies
that this model is not appropriate. However, a similar spectrum could be given by a smaller
 starburst which is less luminous and therefore has the same dust temperature distribution.
 
 In Paper I, it was argued that the possible reason for the appearance of a silicate feature 
in emission in a type 2 AGN was the dominance of the emission from the NLR 
clouds over the emission from an edge-on smooth torus which has a large opening angle and 
therefore highly anisotropic emission
(Efstathiou \& Rowan-Robinson 1995). Another effect that  was first noted by Efstathiou, 
Hough \& Young (1995) who modelled the nuclear IR spectrum of NGC 1068 is that if the 
central source of radiation is an accretion disc, the radiation directed towards the NLR 
 may be a factor of a few more intense than the radiation directed towards the bulk of the 
torus. This effect was also discussed more recently by Kawaguchi \& Mori (2010) who argued that
 the reason that the inner radius of the torus is systematically about a factor of 3 smaller
 than predicted from the AGN luminosity is due to the fact that the central source is an accretion
 disc.
 
The suggestion that the silicate emission feature in a type 2 AGN  is due to emission by NLR
  clouds also seemed to explain the fact that the average mid-IR spectrum of X-ray-selected
 QSO2s also showed a silicate emission feature (Sturm et al. 2006). Since then the same effect
 has been observed in a few Seyfert 2 galaxies, e.g. NGC 2110 (Mason et al. 2009) and a type 2 quasar
 (Nikutta et al. 2009). However, Martinez-Sansigre et al. (2008) and Hernan-Caballero et al. (2009) 
found no evidence for this effect in samples of luminous type 2 AGNs observed with the IRS onboard 
{\em Spitzer}. These results suggest that the deduced covering factor of the NLR in IRAS F10214+4724 (17\% 
or more if the 100K emission is also from the NLR) is either unusually high or the unique spectrum is 
due to the preferential magnification of the emission of the NLR clouds. Mor et al. (2009)  estimate 
that the average NLR covering factor in a sample of Palomar-Green (PG) quasars is 7\%, with values as large as 42\% in
 the case of PG 0157+001. Deane (2012) and Deane et al. (2013) have given support to the idea of 
preferential magnification in IRAS F10214+4724.

 What is also puzzling is the fact that the observed spectrum requires the NLR dust
 to be concentrated in clouds with three discrete values
of dust temperature instead of a continuous distribution of temperature. This may be evidence for 
episodic outflows of material that forms dust. Lacy, Rawlings \& Serjeant (1998) also suggested on the basis of 
near-IR and optical spectroscopy that NLR clouds start life close to the nucleus
 and flow out. Alternatively, this may suggest that it is mainly the emission of the 100-200 K clouds
 that is preferentially magnified.

 In conclusion, IRAS F10214+4724 is probably both a hidden quasar and a protogalaxy in the sense that
 a massive stellar population is in the process of formation at high redshift. We now find, in contrast
 to the predictions of the radiative transfer models of the 1990s, that the bulk of the apparent luminosity
 originates from the NLR.  The model that came closer to the picture we now have is the 
`anisotropic sphere' model of Rowan-Robinson et al. (1993) which was described in more detail by Efstathiou 
\& Rowan-Robinson (1995). The model predicted a silicate emission feature, but its predicted near-IR continuum 
was too strong. The case of IRAS F10214+4724 illustrates the need for rest-frame mid-IR spectroscopy for
 understanding the high-redshift galaxy populations that have been uncovered by galaxy surveys at IR and 
submillimetre wavelengths. The case of IRAS F10214+4724 also shows that it is not at all straightforward 
to interpret the SEDs of objects that are gravitationally lensed. Well-sampled SEDs are needed for understanding 
the emission mechanisms involved and determining the fractions of luminosity that are contributed by the different 
components.

\begin{figure}
\epsfig{file=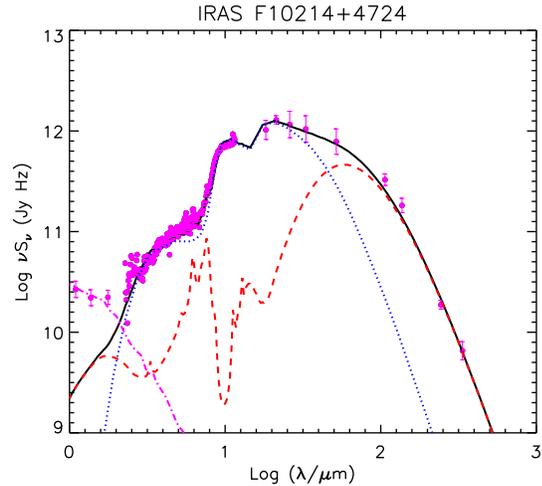, angle=0, width=8cm}
\caption{
Fit to the SED of IRAS F10214+4724. The dotted (blue) and dashed (red) lines
 show the contributions of the AGN and starburst, respectively, to the total emission (solid line). The 
dot-dashed  (magenta) line also shows a fit to the IRAC data with a  2.5-Gyr-old stellar population with 
a luminosity of $6.5 \times 10^{12} L_\odot$. References for the data are given in Table 1.
}
\end{figure}

\begin{figure}
\epsfig{file=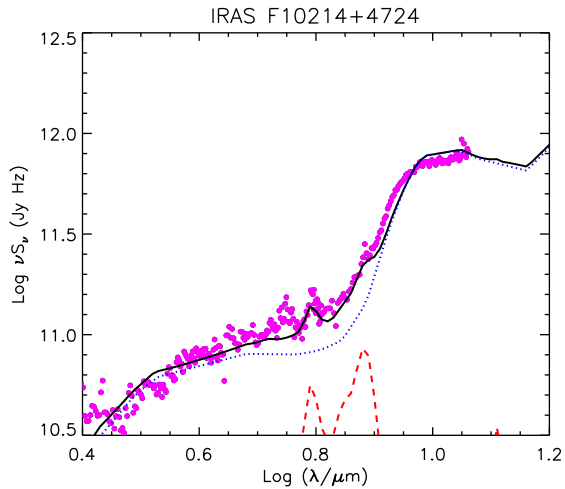, angle=0, width=8cm}
\caption{
As Fig. 6 but showing the fit to the mid-IR spectrum in more detail.
}
\end{figure}

\begin{figure}
\epsfig{file=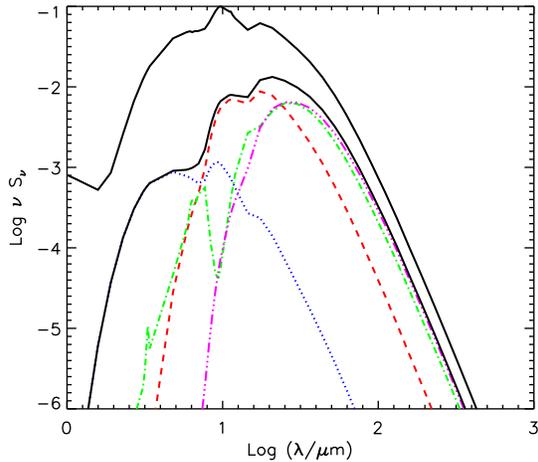, angle=0, width=8cm}
\caption{
Spectra of the three NLR components: 610K (blue dotted line), 200K (red dashed line) 
and 100K (magenta triple dot-dashed line). Also shown in green (dot-dashed line) is the spectrum of
 the `hot' starburst discussed in Section 5. The sum of the emission from the 610, 200 and 100K components
 gives the total emission from the AGN component (black solid line) which is used in the fits shown in Figs 6 
and 7. Also shown with the upper black solid curve is the predicted spectrum of a shell of dust with a 
continuous distribution of dust and an inner dust temperature of 610K (see text for more details).
}
\end{figure}

\section*{Acknowledgements}

 AE acknowledges the hospitality of the Astrophysics group at Oxford during his visit in 2010 July
 where this work began. The authors also benefited from very useful discussions with Steve Rawlings
 and Roger Deane. We would also like to thank an anonymous referee for useful comments that led to an 
improvement of this paper.

\end{document}